\pdfoutput=1
\documentclass[11pt, a4paper]{article}
\usepackage{relsize}
\usepackage{array}
\usepackage{a4wide}
\usepackage[hidelinks]{hyperref}
\usepackage{latexsym}
\usepackage{cite}
\usepackage{mathrsfs}
\usepackage{amssymb}
\usepackage{amsmath}
\usepackage{amsfonts}
\usepackage{amsthm}
\usepackage{graphicx}
\usepackage{enumerate}
\usepackage[usenames,dvipsnames]{xcolor}
\usepackage{todonotes}
\usepackage{bm}

\usepackage{float}
\usepackage{mathtools}
\usepackage{subcaption}
\usepackage{xcolor}

\graphicspath{ {Images/} }

\renewcommand{\vec}{\bs}  
\newcommand{\bs}[1]{\boldsymbol{#1}}

\usepackage{etoolbox}
\apptocmd{\sloppy}{\hbadness 10000\relax}{}{}

\newcommand{\be}{\begin{equation}\label}
\newcommand{\ee}{\end{equation}}
\newcommand{\bea}{\begin{eqnarray}\label}
\newcommand{\eea}{\end{eqnarray}}

\newcommand{\la}{\langle}
\newcommand{\ra}{\rangle}

\makeatletter
\newcommand*{\textoverline}[1]{$\overline{\hbox{#1}}\m@th$}
\newcommand*\bigcdot{\mathpalette\bigcdot@{.65}}
\newcommand*\bigcdot@[2]{\mathbin{\vcenter{\hbox{\scalebox{#2}{$\m@th#1\bullet$}}}}}
\makeatother

\newcommand{\ang}[1]{\left\langle #1\right\rangle}

\newcommand{\angb}[2]{\left\langle #1\bar{#2}\right\rangle}
\newcommand{\angbb}[2]{\left\langle \bar{#1}\bar{#2}\right\rangle}

\newcommand{\ltilde}[1]{\widetilde{\lambda}_{#1}}
\newcommand{\lbar}[1]{\overline{\lambda}_{#1}}

\newcommand{\eq}[1]{\begin{equation}#1\end{equation}}
\newcommand{\eqs}[1]{\begin{equation}\begin{split}#1\end{split}\end{equation}}

\newcommand{\modks}{k_{\underline{12}}}
\newcommand{\modkt}{k_{\underline{23}}}
\newcommand{\modku}{k_{\underline{13}}}

\date{}

\begin{document}

\title{Color/Kinematics Duality in AdS$_4$}

\author{Connor Armstrong, Arthur E. Lipstein and Jiajie Mei \vspace{7pt}\\ \normalsize \textit{
Department of Mathematical Sciences}\\\normalsize\textit{Durham University, Durham, DH1 3LE, United Kingdom}}

{\let\newpage\relax\maketitle}
\begin{abstract}
In flat space, the color/kinematics duality states that perturbative Yang-Mills amplitudes can be written in such a way that kinematic numerators obey the same Jacobi relations as their color factors. This remarkable duality implies BCJ relations for Yang-Mills amplitudes and underlies the double copy to gravitational amplitudes. In this paper, we find analogous relations for Yang-Mills amplitudes in AdS$_4$. In particular we show that the kinematic numerators of 4-point Yang-Mills amplitudes computed via Witten diagrams in momentum space enjoy a generalised gauge symmetry which can be used to enforce the kinematic Jacobi relation away from the flat space limit, and we derive deformed BCJ relations which reduce to the standard ones in the flat space limit. We illustrate these results using compact new expressions for 4-point Yang-Mills amplitudes in AdS$_4$ and their kinematic numerators in terms of spinors. We also spell out the relation to 3d conformal correlators in momentum space, and speculate on the double copy to graviton amplitudes in AdS$_4$.
\end{abstract}

\pagebreak
\tableofcontents

\section{Introduction}

Scattering amplitudes and correlation functions are the basic observables of quantum field theory and in recent years there has been increasing interest in understanding their relation, especially for correlators in conformal field theories which are of immense interest for the study of quantum gravity via the AdS/CFT correspondence \cite{Maldacena:1997re}. In particular, conformal correlators reduce to scattering amplitudes in one  dimension higher when a certain limit is taken corresponding to the flat space limit in the bulk \cite{Penedones:2010ue, Raju:2012zr}. This raises the exciting prospect of adapting some of the powerful techniques for computing scattering amplitudes in flat space to compute conformal correlators, or equivalently amplitudes in AdS.

In this paper, we will consider the generalisation of a remarkable property of perturbative scattering amplitudes known as color-kinematics  (CK) duality. This states that gauge theory amplitudes can be written in such a way that kinematic numerators obey relations analogous to Jacobi relations for their color factors \cite{Bern:2008qj}. Using this decomposition, it is then possible to obtain gravitational amplitudes by replacing color factors with another set of kinematic numerators, implying a general relation between gauge and gravitational scattering amplitudes known as the double copy, which was first seen in the context string amplitudes in the form of the KLT relations \cite{Kawai:1985xq}. Another important implication of the CK duality is a set of relations among gauge theory amplitudes known as the BCJ relations, generalising a 4-point result first discovered in \cite{Zhu:1980sz}. The tree-level double copy was proven from various points of view in \cite{BjerrumBohr:2009rd,Stieberger:2009hq,Bern:2010yg,Feng:2010my} and there is also considerable evidence that it extends to loop level \cite{Bern:2010ue, Carrasco:2011mn, Bern:2012uf}. For a recent review see \cite{Bern:2019prr}.

To adapt these ideas to AdS, we will work with Witten diagrams in momentum space \cite{Raju:2010by,Maldacena:2011nz,Raju:2011mp,Raju:2012zr,Raju:2012zs,Albayrak:2018tam,Albayrak:2019yve}. As explained above, these encode conformal correlators in the boundary. General solutions to the conformal Ward identities in momentum space at three-points were obtained in \cite{Bzowski:2013sza,Bzowski:2017poo,Bzowski:2018fql}, and a generalisation to any number of points for scalar correlators was recently found in \cite{Bzowski:2019kwd}. The study of conformal correlators in momentum space is also relevant to inflationary cosmology  \cite{Maldacena:2011nz,Bzowski:2012ih,Arkani-Hamed:2015bza,Arkani-Hamed:2018kmz,Baumann:2020dch}. More recently, there has also been progress in computing cosmological observables in Mellin space \cite{Sleight:2019mgd}.

At three-points, the double copy amounts to simply squaring gauge theory amplitudes into gravitational ones. The analogous statement for conformal correlators in momentum space is more intricate and was worked out in \cite{Farrow:2018yni,Lipstein:2019mpu}. In particular these references demonstrated that general solutions of the conformal Ward identities involving currents and stress tensors can be related via a double copy in the flat space limit, and this relation can be extended beyond the flat space limit in three dimensions (i.e. a 4d bulk). In even dimensions, the analysis is more complicated because the correlators exhibit branch cuts and divergences which need to be renormalised. In order to extend this story to 4-point correlators, we must first understand how CK duality is realised. To simplify the  problem, we focus on Yang-Mills (YM) theory in AdS$_4$ rather than current correlators of a generic CFT in the boundary. Using momentum space Witten diagrams, we find that tree-level 4-point amplitudes can be written in terms of kinematic numerators analogous to those of flat space, although these numerators are far more complicated for generic polarisations. Moreover, we find that these numerators do not generically obey a kinematic Jacobi relation away from the flat space limit. On the other hand, the 4-point AdS$_4$ amplitudes enjoy a generalised gauge symmetry which can be used to obtain a unique set of kinematic numerators which obey the Jacobi identity away from the flat space limit. It is then natural to speculate that squaring these preferred numerators gives rise to something closely related to a 4-point graviton amplitude in AdS$_4$, or equivalently a stress tensor correlator in the boundary. Along the way, we also derive deformed BCJ relations in AdS$_4$ which reduce to the standard ones in the flat space limit.

The expressions we obtain for AdS$_4$ amplitudes and their kinematic numerators can be greatly simplified using a spinor-helicity formalism adapted to AdS$_4$ \cite{Maldacena:2011nz,Raju:2012zs,Farrow:2018yni,Baumann:2020dch}. Using numerous spinor identities, we obtain a new formula for the amplitude with two negative and two positive helicity gluons, which is much more compact than the one first obtained in the pioneering work \cite{Raju:2012zs}. We also obtain very concise new formulas for all other helicity configurations which to our knowledge have not been previously computed.

This paper is organised as follows. In section \ref{overview}, we review the CK duality, BCJ relations, and double copy for flat space Yang-Mills amplitudes at 4-points, and describe the generalisation to AdS$_4$. In section \ref{generalpolarization} we derive explicit formulae for 4-point YM amplitudes in AdS$_4$ and their kinematic numerators, showing that they generically do not obey the kinematic Jacobi relation away from the flat space limit. We then derive expressions for a unique choice of kinematic numerators for which the CK duality holds away from the flat space limit. In section \ref{helicityamp} we convert the results of section \ref{generalpolarization} to spinor notation, obtaining very concise new formulas for all helicity configurations. In section \ref{sec:Reconstruction}, we explain how to reconstruct 3d conformal correlators in momentum space  from AdS$_4$ amplitudes by replacing polarisations with transverse projection tensors and adding longitudinal terms fixed by the transverse Ward identities. In section \ref{conclusion}, we speculate on how to obtain 4-point gravitational amplitudes in AdS$_4$ by squaring the preferred kinematic numerators appearing in YM amplitudes, and deduce a deformed KLT relation which reduces to the standard one in the flat space limit. There are also two Appendices. In Appendix \ref{app:wittdiag}, we provide additional details of Witten diagram calculations, and in Appendix \ref{app:spinors}, we derive numerous spinor identities and explain how they are used to simplify Witten diagrams.

\section{Overview} \label{overview}
In this section, we will review some aspects of the CK duality in flat space and generalise them to AdS.

\subsection{Flat space}
Let us begin by reviewing CK duality for tree-level 4-point scattering amplitudes \cite{Bern:2008qj,Bern:2019prr}.  A 4-point color-dressed gluon amplitude can be written as
\begin{align}
    \mathcal{A}_4=\frac{n_s c_s}{s} +\frac{n_t c_t}{t} +\frac{n_u c_u}{u},
\label{colordressed}
\end{align}
where $s,t,u$ are Mandelstam variables, $c_i$ are color factors, $n_{i}$ are kinematic numerators, and we have set the YM coupling to one. The $c_i$ can be written in terms of color group structure constants
\eq{
c_{s}=f^{a_{1}a_{2}b}f^{ba_{3}a_{4}},\,\,\,c_{t}=f^{a_{1}a_{4}b}f^{ba_{2}a_{3}},\,\,\,c_{u}=f^{a_{3}a_{1}b}f^{ba_{2}a_{4}},
\label{eqn:colorFactors}
}
which satisfy the Jacobi relation
\begin{equation}
c_{s}+c_{u}+c_{t}=0.
\label{jacobic}
\end{equation}
Using this to express $c_t$ in terms of $c_s$ and $c_u$ and plugging it back into \eqref{colordressed} gives
\begin{equation}
\mathcal{A}_{4}=c_{s} A_{1234}-c_{u} A_{1342},
\end{equation}
where the color-ordered amplitudes are given by
\eqs{
A_{1234}&=\frac{n_{s}}{s}-\frac{n_{t}}{t},\\
A_{1324}&=\frac{n_{t}}{t}-\frac{n_{u}}{u}.
\label{eq:ymamp}
}

\begin{figure}
\centering
\begin{subfigure}{.33\textwidth}
  \centering
  \includegraphics[width=.9\linewidth]{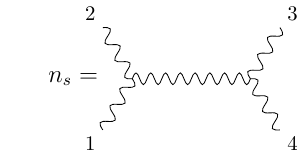}
\end{subfigure}%
\begin{subfigure}{.33\textwidth}
  \centering
  \includegraphics[width=.9\linewidth]{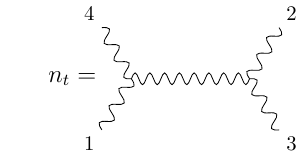}
\end{subfigure}
\begin{subfigure}{.33\textwidth}
  \centering
  \includegraphics[width=.9\linewidth]{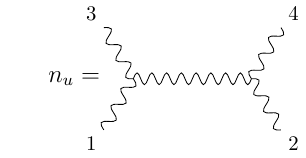}
\end{subfigure}
\caption{The three numerator structures for the color-dressed amplitude. Two of these appear in each color-ordered amplitude. }
\label{fig:numerators}
\end{figure}

The kinematic numerators can be represented using the cubic diagrams in Figure \ref{fig:numerators}. These are not Feynman diagrams, but are derived by splitting the contact Feynman diagram into three pieces multiplied by $s/s$, $t/t$, and $u/u$, respectively, and combining them with exchange Feynman diagrams.
When written in terms of polarisation vectors, the numerators are related by cyclic permutations on three particles:
\eq{
n_t = n_s\Big|_{(234) \to (423)},\qquad n_u = n_s\Big|_{(234) \to (342)},
\label{numeratorrelation}
}
or equivalently using exchanges:
\eq{
n_t = -n_s\big|_{2\leftrightarrow4},\qquad n_u = -n_s\big|_{2\leftrightarrow3}.
\label{eqn:numeratorexchange}
}
These operations can be seen by permuting the labels on the diagrams in Figure \ref{fig:numerators}. When we express the numerators using spinors, these relations continue to hold except when these mappings act on particles of different helicities. Note that exchanging the legs at a vertex gives rise to a minus sign, which can be seen from the antisymmetric structure of the three point vertex. This explains the minus signs in \eqref{numeratorrelation} and the relative signs in \eqref{eq:ymamp}; we get a minus sign when we flip the legs at a vertex so that the ordering of the diagram matches the ordering of the partial amplitude.

The kinematic Jacobi relation states that
\begin{equation}
n_{s}+n_{t}+n_{u}=0,\label{eq:kinjacobi}
\end{equation}
which is analogous to the color Jacobi relation in \eqref{jacobic} and encodes a remarkable duality between color
and kinematics. Given a set of numerators satisfying \eqref{eq:kinjacobi},
one can then construct gravitational amplitudes from color-dressed YM amplitudes by replacing the color factors with kinematic numerators:
\begin{equation}
\mathcal{M}_{4}=\frac{n_{s}^{2}}{s}+\frac{n_{t}^{2}}{t}+\frac{n_{u}^{2}}{u},
\label{gravdouble}
\end{equation}
where we have set the gravitational coupling constant to 1. This relation is referred to as the double copy. The kinematic Jacobi relation can also be used to prove a relation among color-ordered amplitudes known
as the BCJ relation:
\eq{
uA_{1324} = sA_{1234}.
\label{bcjflat}
}
Plugging \eqref{eq:ymamp} and \eqref{bcjflat} into \eqref{gravdouble} then implies the famous KLT relation
\begin{equation}
\mathcal{M}_{4}=-sA_{1234}A_{1243},
\label{kltrelat}
\end{equation}
first discovered in the context of string theory \cite{Kawai:1985xq}.

\subsection{AdS}

The goal of this paper will be to extend the above story to AdS$_4$. In this case, the analogue of Feynman diagrams are Witten diagrams in momentum space, which were previously studied for Yang-Mills theory in \cite{Albayrak:2018tam}. The quantities we obtain from adding up Witten diagrams, which we refer to as AdS amplitudes, encode correlators of conserved currents in a 3d CFT living in the boundary, so we will denote them as $\left\langle jjjj\right\rangle$. In more detail, they encode the transverse part of 3d conformal correlators, from which the full correlators can be reconstructed using Ward identities, as we spell out in section \ref{sec:Reconstruction}. The conformal correlators can be written in terms of 3-momenta $\vec{k}=\left\{ k^{0},k^{1},k^{2}\right\}$, whose sum over all external particles is conserved due to the translational symmetry of the boundary. To relate these to Yang-Mills amplitudes in the bulk, it is convenient to lift the 3-momenta to null 4-momenta as follows:
\eqs{
k^{\mu} &= (k^0,k^1,k^2,ik),\\
\mathrm{where}\quad k &=\left|\vec{k}\right|= \sqrt{-(k^0)^2+(k^1)^2+(k^2)^2},
\label{eqn:nullMomDefn}
}
such that $i k$ is the radial component of the momentum. If $\vec{k}$ is time-like, then $k$ is imaginary but if $\vec{k}$ is space-like then $k$ is real. We will work with space-like boundary momenta. The sum over radial momenta will not vanish in general and we define it as
\eq{
E=\sum_{i=1}^{4}k_{i},
\label{sumr}
}
where $i$ is an external particle label. In the limit $E \rightarrow 0$, 3d correlators develop a pole whose residue is a 4d scattering amplitude in flat space \cite{Raju:2012zr}:
\begin{equation}
\lim_{E\rightarrow0}\left\langle jjjj\right\rangle =\frac{A_{4}}{E}.
\label{flatspacelimit}
\end{equation}

To make the relation to amplitudes more explicit, it is convenient to dress the correlators with polarisations of the form
\begin{equation}
\epsilon^{\mu}=\left(\vec{\epsilon},0\right),
\label{polarisations}
\end{equation}
which satisfy
\begin{equation}
\vec{\epsilon}_{i}\cdot\vec{k}_{i}=0,\,\,\,\vec{\epsilon}_{i}\cdot\vec{\epsilon}_{i}=0.
\label{polarisationconstraints}
\end{equation}
These polarisations also arise when computing Witten diagrams in axial gauge. Since the polarisations are transverse to the momenta, they project out the longitudinal parts of correlators, giving AdS amplitudes. In AdS it is also natural to define the following analogue of Mandelstam variables \cite{Baumann:2020dch}:
\eqs{
s&=\left(k_{12}+k_{\underline{12}}\right)\left(k_{34}+k_{\underline{34}}\right),\\
t&=\left(k_{14}+k_{\underline{14}}\right)\left(k_{23}+k_{\underline{23}}\right),\\
u&=\left(k_{13}+k_{\underline{13}}\right)\left(k_{24}+k_{\underline{24}}\right),
\label{eqn:GenMandelstam}
}
where $k_{\underline{ij}}=\left|\vec{k}_{i}+\vec{k}_{j}\right|$ and
$k_{ij}=k_{i}+k_{j}$. Unlike in flat space, these variables do not
add to zero for massless external states. Their sum is given by
\eqs{
s+t+u &= \xi,\\
\mathrm{where}\quad \xi = E \big(E+k_{\underline{12}}&+k_{\underline{23}}+k_{\underline{13}}\big).
\label{stuxi}
}
In the flat space limit, $\xi\rightarrow 0$ and the Mandelstam variables reduce to their standard definitions after using 4-momentum conservation. Note that our treatment of the double copy in AdS will have some similarities to the double copy of massive amplitudes in flat space, where one also has $s+t+u \neq 0$ \cite{Momeni:2020vvr,Johnson:2020pny}.

Using Witten diagrams, one finds that color-ordered YM amplitudes in AdS$_4$
can also be written in the form \eqref{eq:ymamp} using the generalised
Mandelstam invariants in \eqref{eqn:GenMandelstam}:
\eqs{
\left\langle j_{1}j_{2}j_{3}j_{4}\right\rangle &=\frac{n_{s}}{s}-\frac{n_{t}}{t},\\
\left\langle j_{1}j_{3}j_{2}j_{4}\right\rangle &=\frac{n_{t}}{t}-\frac{n_{u}}{u},\label{eq:ymampads}
}
where the kinematic numerators once again obey \eqref{numeratorrelation} and \eqref{eqn:numeratorexchange}. Using these relations among kinematic numerators we can also show that color-ordered AdS amplitudes obey a photon decoupling relation analogous to that of flat space amplitudes:
\eq{
\ang{j_1j_2j_3j_4} + \ang{j_1j_3j_4j_2} + \ang{j_1j_4j_2j_3} = 0.
\label{eqn:photondecoupling}
}
Unlike in flat space, however, in AdS the kinematic numerators do not generically add
to zero, and we will denote their sum as $Q$:
\begin{equation}
Q=n_{s}+n_{t}+n_{u}.\label{eq:adsq}
\end{equation}
Using this equation, we can eliminate $n_{t}$ in \eqref{eq:ymampads}
to obtain the following relation between the color-ordered AdS amplitudes
and kinematic numerators:
\eq{
\left(\begin{array}{c}
\left\langle j_{1}j_{2}j_{3}j_{4}\right\rangle + Q/t\\
\left\langle j_{1}j_{3}j_{2}j_{4}\right\rangle - Q/t
\end{array}\right)=\left(\begin{array}{cc}
1/s+1/t & 1/t\\
-1/t & -1/u-1/t
\end{array}\right)\left(\begin{array}{c}
n_{s}\\
n_{u}
\end{array}\right).
}
In flat space, the matrix on the right is not invertible, which is expected
since amplitudes on the left are gauge invariant while numerators
on the right are not. In AdS however, we can invert the matrix to
obtain
\eq{
\left(\begin{array}{c}
n_{s}\\
n_{u}
\end{array}\right)=\frac{1}{\xi}\left(\begin{array}{cc}
s(t+u) & su\\
-su & -u(s+t)
\end{array}\right)\left(\begin{array}{c}
\left\langle j_{1}j_{2}j_{3}j_{4}\right\rangle + Q/t\\
\left\langle j_{1}j_{3}j_{2}j_{4}\right\rangle - Q/t
\end{array}\right).
\label{numfromjjjj}
}
Note that the inverse becomes singular in the flat space limit as $\xi\rightarrow0$.
The solution for $n_{u}$ implies
\eq{
u\left\langle j_{1}j_{3}j_{2}j_{4}\right\rangle - s\left\langle j_{1}j_{2}j_{3}j_{4}\right\rangle=\xi\left(\left\langle j_{1}j_{3}j_{2}j_{4}\right\rangle+\frac{n_{u}}{u}\right)-Q.
}
Plugging (\ref{eq:ymampads}) into the right hand side finally gives
the deformed BCJ relation:
\begin{equation}
u\left\langle j_{1}j_{3}j_{2}j_{4}\right\rangle-s\left\langle j_{1}j_{2}j_{3}j_{4}\right\rangle=\xi\frac{n_{t}}{t}-Q.\label{bcjads}
\end{equation}
In the flat space limit, which is defined by multiplying by $E$ and taking $E \rightarrow 0$, it is not difficult to see that this reduces to the standard BCJ relation in \eqref{bcjflat}.

Note the AdS amplitudes in \eqref{eq:ymamp} are invariant under
the following generalised gauge transformation of the kinematic numerators:
\begin{equation}
n_{s}\rightarrow n_{s}+s\Delta,\,\,\,n_{t}\rightarrow n_{t}+t\Delta,\,\,\,n_{u}\rightarrow n_{u}+u\Delta,
\label{gauge}
\end{equation}
where $\Delta$ is an arbitrary function. Under this transformation, the parameter $Q$ in \eqref{eq:adsq} transforms as
\eq{
Q\rightarrow Q+\xi\Delta.
}
Hence, by choosing $\Delta=-Q/\xi$, we can set $Q$ to zero. We shall denote the kinematic numerators in this generalised gauge as $\tilde{n}_{s}$,
$\tilde{n}_{t}$, $\tilde{n}_{u}:$
\begin{equation}
\tilde{n}_{s}=n_{s}-sQ/\xi,\,\,\,\tilde{n}_{t}=n_{t}-tQ/\xi,\,\,\,\tilde{n}_{u}=n_{u}-uQ/\xi.
\label{ntilde}
\end{equation}
In this generalised gauge, the numerators satisfy the kinematic Jacobi relation away from the flat space limit:
\eq{
\tilde{n}_{s}+\tilde{n}_{t}+\tilde{n}_{u}=0.
\label{adskinematicjacobi}
}
Hence, these numerators can also be obtained from \eqref{numfromjjjj} by setting $Q=0$ on the right hand side. Moreover, the deformed BCJ relation in \eqref{bcjads} reduces to
\eq{
u\left\langle j_{1}j_{3}j_{2}j_{4}\right\rangle-s\left\langle j_{1}j_{2}j_{3}j_{4}\right\rangle=\xi\frac{\tilde{n}_{t}}{t}.
\label{deformedbcj2}
}
In flat space the generalised gauge transformations in \eqref{gauge} do not affect the sum of kinematic numerators and correspond to ordinary gauge transformations for a particular choice of $\Delta$. In the next section, we will derive explicit formulas for 4-point AdS amplitudes and their kinematic numerators using Witten diagrams.

\section{AdS$_4$ Amplitudes with General Polarizations} \label{generalpolarization}

In this section, we will derive formulas for 4-point Yang-Mills amplitudes in AdS$_4$ using Witten diagrams in momentum space.  These were previously computed in \cite{Albayrak:2018tam}, but here we will make the results more explicit and use them to derive formulae for numerators obeying kinematic Jacobi identities

\begin{figure}
\centering
\includegraphics[width=15cm]{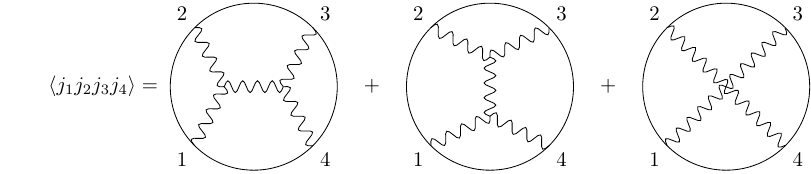}
\caption{Witten diagrams for the color-ordered 4-point AdS amplitude.}
\label{fig:WittenDiag4ptAmp}
\end{figure}
For color-ordered amplitudes, there are three Witten diagrams, which are shown in Figure \ref{fig:WittenDiag4ptAmp}. We will denote the contact diagram as $W_c$ and the two exchange diagrams as $W_s$ and $W_t$. The $s$-channel diagram is given by
\eq{
	W_s=-\frac{1}{E s} \left(V^{12m} V^{34}_m+ \frac{E 	+k_{\underline{12}}}{k_{12}k_{34}k_{\underline{12}}}V^{12m} (\vec{k}_{12})_mV^{34n} (\vec{k}_{12})_n\right),
\label{schannel}
}
where $V_{k}^{12}=\epsilon_{1}^{i}\epsilon_{2}^{j}V_{ijk}$ are 3-point vertices dressed with polarisation vectors, and the index $k=0,1,2$ denotes a boundary direction. The 3-point vertices take the same form as in flat space but with components restricted to boundary directions and are given in \eqref{ymvertices}. For a derivation of \eqref{schannel}, see Appendix \ref{app:wittdiag}. In the flat-space limit, this reduces to the usual Feynman diagram expression in axial gauge:
\eq{
\lim_{E\to 0}E W_s =V^{12m}V_{34m}\frac{1}{(k_{1\mu}+k_{2\mu})^2}\left(-\eta _{mn}+\frac{\vec{k}_{12m}\vec{k}_{12n}}{k_{12}^2}\right).
}
Writing out the 3-point vertices and using momentum conservation then gives
\eq{
    W_s=\frac{1}{2}\frac{\epsilon _1 \!\cdot\! \epsilon _2\, \epsilon _3\!\cdot\! \epsilon _4}{E s}[(k_{1\mu}-k_{2\mu})(k_3^{\mu}-k_4^{\mu})-\frac{E}{k_{\underline{12}}}(k_1-k_2)(k_3-k_4)]+ W^B_{s},
\label{eqn:sChannelWittenDiag}
}
where inner products of 4-vectors are taken with respect to a flat metric, and $W^B_{s}$ consists of the terms not proportional to $\epsilon_1 \!\cdot\! \epsilon_2\, \epsilon_3\!\cdot\! \epsilon_4$:
\begin{align}
    W_s^B= \frac{1}{E} \frac{1}{s} \big[&\epsilon _1\cdot \epsilon _2 \boldsymbol{k} _4 \cdot \epsilon _3 (\boldsymbol{k}_1-\boldsymbol{k}_2)\cdot \epsilon _4-\epsilon _1\cdot \epsilon _2 \boldsymbol{k} _3 \cdot \epsilon _4 (\boldsymbol{k}_1-\boldsymbol{k}_2)\cdot \epsilon _3 \nonumber\\
    &+\epsilon _3\cdot \epsilon _4 \boldsymbol{k} _2 \cdot \epsilon _1 (\boldsymbol{k}_3-\boldsymbol{k}_4)\cdot \epsilon _2-\epsilon _3\cdot \epsilon _4 \boldsymbol{k} _1 \cdot \epsilon _2 (\boldsymbol{k}_3-\boldsymbol{k}_4)\cdot \epsilon _1 \nonumber\\
    &+2\epsilon _2\cdot \epsilon _4 \boldsymbol{k} _2 \cdot \epsilon _1 \boldsymbol{k}_4\cdot \epsilon _3-2\epsilon _2\cdot \epsilon _3 \boldsymbol{k} _2 \cdot \epsilon _1 \boldsymbol{k}_3\cdot \epsilon _4 \nonumber \\
    &-2\epsilon _1\cdot \epsilon _4 \boldsymbol{k} _1 \cdot \epsilon _2 \boldsymbol{k}_4\cdot \epsilon _3 +2\epsilon _1\cdot \epsilon _3 \boldsymbol{k} _1 \cdot \epsilon _2 \boldsymbol{k}_3\cdot \epsilon _4 \big].
\label{eqn:sChannelBTerm}
\end{align}
The $t$-channel diagram is obtained by swapping $2 \leftrightarrow 4$ in the above formula, and the contact diagram is given by
\begin{align}
    W_c&=\frac{1}{E}\big(\epsilon _1\!\cdot\! \epsilon _3\, \epsilon _2\!\cdot\! \epsilon _4 -\frac{1}{2}(\epsilon _1\!\cdot\! \epsilon _2\, \epsilon _3\!\cdot\! \epsilon _4 +\epsilon _1\!\cdot\! \epsilon _4\, \epsilon _2\!\cdot\! \epsilon _3 )\big).
\end{align}
The full color-ordered AdS$_4$ amplitude is therefore given by
\begin{align}
    \ang{j_1j_2j_3j_4}=&\frac{1}{2E}\frac{\epsilon _1 \!\cdot\! \epsilon _2\, \epsilon _3\!\cdot\! \epsilon _4}{s}[(k_{1\mu}-k_{2\mu})(k_3^{\mu}-k_4^{\mu})-\frac{E}{\modks}(k_1-k_2)(k_3-k_4)]+W^B_{s} \nonumber \\
    &+\frac{1}{2E}\frac{\epsilon _1 \!\cdot\! \epsilon _4\, \epsilon _2\!\cdot\! \epsilon _3}{t}[(k_{1\mu}-k_{4\mu})(k_3^{\mu}-k_2^{\mu})-\frac{E}{\modkt}(k_1-k_4)(k_3-k_2)]+W^B_{t} \nonumber \\
    &+\frac{1}{E}(\epsilon _1\!\cdot\! \epsilon _3\, \epsilon _2\!\cdot\! \epsilon _4 -\frac{1}{2}(\epsilon _1\!\cdot\! \epsilon _2\, \epsilon _3\!\cdot\! \epsilon _4 +\epsilon _1\!\cdot\! \epsilon _4\, \epsilon _2\!\cdot\! \epsilon _3 )).
\label{4ptcolorordered}
\end{align}

Let us now derive kinematic numerators by writing \eqref{4ptcolorordered} in the form \eqref{eq:ymamp}. In doing so, we must split the contact term into two pieces multiplied by $s/s$ and $t/t$, respectively, such that the resulting numerators obey \eqref{numeratorrelation} or equivalently \eqref{eqn:numeratorexchange}. A natural choice for the s-channel numerator is
\begin{align}
   n_s &= \frac{1}{2}\frac{\epsilon _1 \!\cdot\! \epsilon _2\, \epsilon _3\!\cdot\! \epsilon _4}{E}\left[(k_{1\mu}-k_{2\mu})(k_3^{\mu}-k_4^{\mu})-\frac{E}{k_{\underline{12}}}(k_1-k_2)(k_3-k_4)\right] \nonumber \\
   &+ s W^B_s+s\frac{1 }{2E}\left[\epsilon _1 \!\cdot\! \epsilon _3\, \epsilon _2 \!\cdot\! \epsilon _4 - \epsilon _1 \!\cdot\! \epsilon _4\, \epsilon _2 \!\cdot\! \epsilon _3\right].
\label{eqn:sNumeratorAll}
\end{align}
In the flat space limit, this expression reduces to the one obtained in \cite{Bern:2019prr}. The kinematic numerators $n_t$ and $n_u$ can be deduced from \eqref{eqn:sNumeratorAll} using the relations in \eqref{numeratorrelation} or \eqref{eqn:numeratorexchange}. After some algebra, the sum of kinematic numerators is given by
\begin{align}
  Q=\frac{\epsilon _1 \!\cdot\! \epsilon _2\,\epsilon _3 \!\cdot\! \epsilon _4}{2}[(\modkt-\modku)-\frac{1}{\modks}(k_1-k_2)(k_3-k_4)]+\mathrm{Cyc}[234].
\label{eqn:JacobiFailure}
\end{align}
We therefore see that for this choice of numerators, the kinematic Jacobi relation is only satisfied in the flat space limit (recall that taking the flat space limit involves multiplying by $E$ and taking $E \rightarrow 0$). However in AdS$_4$ we can use the generalised gauge symmetry in \eqref{gauge} to obtain numerators that obey the kinematic Jacobi relation even away from the flat space limit. The preferred numerators $\left\{ \tilde{n}_{s},\tilde{n}_{t},\tilde{n}_{u}\right\} $ are obtained by plugging \eqref{eqn:sNumeratorAll} and \eqref{eqn:JacobiFailure} into \eqref{ntilde} and using \eqref{eqn:numeratorexchange}.

We can prove \eqref{eqn:JacobiFailure} by summing the kinematic numerators and first considering the terms containing four polarisation vectors contracted together. For example, we get the following terms proportional to $\epsilon _1 \!\cdot\! \epsilon _2\, \epsilon _3\!\cdot\! \epsilon _4$:
\begin{align}
    \epsilon _1 \!\cdot\! \epsilon _2\, \epsilon _3\!\cdot\! \epsilon _4[(k_{1\mu}-k_{2\mu})(k_{3\mu}-k_{4\mu})+t-u]=\epsilon _1 \!\cdot\! \epsilon _2\, \epsilon _3\!\cdot\! \epsilon _4 [E(\modkt-\modku)],
\label{eqn:Q_Proof}
\end{align}
where we have dropped for now the terms with a pole in $k_{\underline{ij}}$.
From the $W^B$ terms appearing in the kinematic numerators, let us consider the term proportional to $\epsilon _2 \cdot \epsilon _3$:
\begin{align}
    2\epsilon _2\!\cdot\! \epsilon _3\, \boldsymbol{k} _2 \!\cdot\! \epsilon _1\, \boldsymbol{k}_3\!\cdot\! \epsilon _4 + 2\epsilon _2\!\cdot\! \epsilon _3\, (\boldsymbol{k} _1 \!\cdot\! \epsilon _4\, \boldsymbol{k}_2\!\cdot\! \epsilon _1-\boldsymbol{k} _4 \!\cdot\! \epsilon _1\, \boldsymbol{k}_2\cdot \epsilon _4)-2\epsilon _2\!\cdot\! \epsilon _3\, \boldsymbol{k} _3 \!\cdot\! \epsilon _1\, \boldsymbol{k}_2\cdot \epsilon _4=0,
\end{align}
where the first term comes from $n_s$, the second term from $n_t$, and the third term from $n_u$. All the contributions from $W_B$ drop out in an analogous manner. We are therefore only left with the terms in equation \eqref{eqn:Q_Proof} and the $k_{\underline{ij}}$ pole, which sum to \eqref{eqn:JacobiFailure}.\\

\section{AdS$_4$ Helicity Amplitudes} \label{helicityamp}

In this section, we will specialise the results obtained in the previous section to particular helicities of the external gluons and write them in terms of spinors adapted to AdS$_4$. The AdS$_4$ amplitudes will then be labelled by the helicities of the external gluons $\left\langle h_{1}h_{2}h_{3}h_{4}\right\rangle $, where $h_{i}=\pm$. An amplitude with $k$ negative helicity gluons is referred to as an N$^{k-2}$MHV amplitude. In the flat space limit, only the $k=2$ (or MHV) amplitude is non-zero at tree-level but in AdS$_4$, the amplitude is non-zero for $k=0,1$ as well. Amplitudes with $k=3,4$ are related to $k=1,0$ via parity. A spinorial expression for the MHV amplitude was previously obtained in \cite{Raju:2012zs} using a recursive approach. Using numerous identities derived in Appendix \ref{app:spinors}, we obtain a new expression which appears to be much simpler. We also obtain compact new expressions for non-MHV amplitudes, which to our knowledge were not previously computed.

Let us briefly review the spinor-helicity formalism for AdS$_4$, which can also be found in several other references, although with slightly different conventions \cite{Maldacena:2011nz,Raju:2012zs,Farrow:2018yni,Baumann:2020dch}. For a more detailed discussion and a list of many useful identities, see Appendix \ref{app:useful identities}. The starting point is the bi-spinor form for a null momentum in 4d Minkowski space:
\eq{
k^{\alpha\dot{\alpha}} = k^\mu \sigma_\mu^{\alpha \dot{\alpha}} =\lambda^{\alpha}\ltilde{}^{\dot{\alpha}},
\label{4dspin}
}
where $\sigma_\mu$ are the Pauli matrices, and $\alpha,\dot{\alpha}$ are spinor indices, noting that the Lorentz group is locally $SU(2)\times SU(2)$. Recall that in AdS$_4$, momentum is not conserved along the radial direction, which we denote by the 4-vector $T^\mu=\left\{ 0,0,0,1\right\} $. One can then use $T^{\alpha \dot{\alpha}}=T^{\mu} \sigma_\mu^{\alpha \dot{\alpha}}$ to convert dotted indices to undotted indices, breaking the Lorentz group to its diagonal subgroup corresponding to three dimensional Lorentz group in the boundary \cite{Lipstein:2012kd}. Defining
\eq{
\lbar{i}^\alpha =T^{\alpha\dot{\alpha}}\tilde{\lambda}_{\dot{\alpha}},
}
it becomes possible to take inner products of barred and unbarred spinors:
\eq{
\varepsilon_{\alpha\beta}\lambda_i^\alpha\lbar{j}^\beta = \angb{i}{j} = -\ang{\bar{j}i},
\label{2bracket}
}
where $\varepsilon^{\alpha \beta}$ is the Levi-Civita symbol given by $\varepsilon^{12}  = -\varepsilon_{12} = +1$, and there is a similar definition for the inner product of two barred or two unbarred spinors. A very useful formula is the Schouten identity:
\eq{
\lambda_{i}\left\langle jk\right\rangle +{\rm cyclic}=0,
\label{shouten}
}
which also holds for mixed brackets (so long as bars are permuted along with the particle labels).

After identifying dotted and undotted spinors, the 4d null momentum in \eqref{4dspin} can be written as
\eq{
k^{\alpha\beta}= \lambda^{\alpha}\bar{\lambda}^{\beta}=\lambda^{(\alpha}\bar{\lambda}^{\beta)}+ik\varepsilon^{\alpha\beta},
\label{adsbispin}
}
where we have symmetrised and antisymmetrised the spinor indices. Note that the first term on the right hand side corresponds to the 3-momentum along the boundary, while the second term is the radial momentum. Recalling that momentum along the boundary is conserved, we immediately obtain
\eq{
\sum_{i=1}^n \lambda_i^\alpha\lbar{i}^\beta = i \varepsilon^{\alpha\beta}E,
\label{esum}
}
where $i$ labels external particles and $E$ is defined in \eqref{sumr}. Many other useful identities can then be obtained by contracting \eqref{esum} with various combinations of spinors.

The final ingredient needed to convert Witten diagrams to spinor notation is a formula for the polarisation vectors which are given by:
\eq{
\epsilon^+_{\alpha\beta} = \frac{\lbar{\alpha}\lbar{\beta}}{ik},\qquad \epsilon^-_{\alpha\beta} = \frac{\lambda_\alpha\lambda_\beta}{ik}.
\label{polspin}
}
Note that this definition automatically satisfies \eqref{polarisationconstraints}. Moreover, performing a decomposition analogous to the one in \eqref{adsbispin}, we see that the radial component vanishes in agreement with \eqref{polarisations}.

\subsection{N$^{-2}$MHV}
We will start with the case $++++$. Writing the kinematic numerator in \eqref{eqn:sNumeratorAll} in terms of spinors and using the Schouten identity to rewrite the contact diagram as
\eqs{\frac{1}{2E}\left [\epsilon _1 \!\cdot\! \epsilon _3\, \epsilon _2 \!\cdot\! \epsilon _4 - \epsilon _1 \!\cdot\! \epsilon _4\, \epsilon _2 \!\cdot\! \epsilon _3\right ]=&\frac{1}{8Ek_1k_2k_3k_4}
\left (\angbb{1}{3}^2 \angbb{2}{4}^2 -\angbb{1}{4}^2 \angbb{2}{3}^2 \right ) \\
= &\frac{1}{8Ek_1k_2k_3k_4}\left (2\angbb{1}{3}\angbb{2}{4}\angbb{1}{2}\angbb{3}{4} - \angbb{1}{2}^2\angbb{3}{4}^2\right ),
}
we obtain
\eqs{
n_s^{++++} =& \frac{1}{8k_1k_2k_3k_4}\frac{1}{E} \Big [\angbb{1}{2}^2 \angbb{3}{4}^2
\left(\ang{13}\angbb{1}{3} +\ang{24}\angbb{2}{4} -\frac{E}{k_{\underline{12}}}(k_1-k_2)(k_3-k_4)-E \modks \right)  \\
&\qquad\qquad+2s \angbb{1}{3}\angbb{2}{4}\angbb{1}{2}\angbb{3}{4} \Big ]+s W_s^{B,++++}.
}
where $W_s^{B, ++++}$ has a very compact form given by
\eqs{
W_s^{B, ++++}=\frac{i}{8k_1k_2k_3k_4}\frac{2}{s}\angbb{1}{2}\angbb{3}{4} (\angb{1}{2}\angbb{4}{1}\angbb{1}{3} + \angb{2}{1}\angbb{3}{2}\angbb{2}{4}).
\label{Wbpppp}
}
We explain how \eqref{Wbpppp} was derived in Appendix \ref{app:BTerms} using the $-+++$ case as an example.

To achieve further cancellations, it is convenient to keep the maximum amount of symmetry in our expressions. We therefore write the generalised Mandelstam variable $s$ as
\eq{
s=\frac{1}{2}(\ang{12} \angbb{1}{2}+\ang{34} \angbb{3}{4}+E(E+\modks)).
\label{eqn:genMandelS}
}
We then apply various spinor identities in Appendix \ref{app:useful identities} to terms with a (spurious) $1/E$ pole to get
\eqs{
n_s^{++++} =& \frac{1}{8k_1k_2k_3k_4}\frac{\angb{1}{2} \angbb{3}{4}}{E} \Big [i E(\angb{1}{4}\angbb{1}{2} \angbb{1}{3}+\angb{4}{1}\angbb{3}{4}\angbb{2}{4})  \\
&+\angbb{1}{2}\angbb{3}{4}
\left( -\frac{E}{\modks}(k_1-k_2)(k_3-k_4)-E \modks \right)
+(E^2+2 E\modks) \angbb{1}{3}\angbb{2}{4}\Big ]+s W_s^{B,++++},
}
and then use \eqref{eqn:MB in s and t} to combine the first line with $W_s^{B,++++}$ to obtain the final form
\eqs{
n_s^{++++} &= \frac{1}{8k_1k_2k_3k_4}\angbb{1}{2}\angbb{3}{4}\Big[i\big(\angb{1}{2}\angbb{4}{1}\angbb{1}{3} + \angb{2}{1}\angbb{3}{2}\angbb{2}{4}\big)\\
&\qquad - \modks \left(\angbb{2}{3}\angbb{4}{1} - \angbb{1}{3}\angbb{2}{4}\right)-\frac{1}{\modks}\angbb{1}{2}\angbb{3}{4}(k_1 - k_2)(k_3 - k_4)\Big].
\label{eqn:AllPlusNumerator}
}
The other kinematic numerators can then be obtained using \eqref{eqn:numeratorexchange}. Summing the three numerators and applying the Schouten identity gives
\eqs{
Q^{++++} = \frac{1}{8k_1k_2k_3k_4}\left(\modks \left(\angbb{1}{3}^2\angbb{2}{4}^2 - \angbb{2}{3}^2\angbb{4}{1}^2 \right) - \frac{1}{\modks}\angbb{1}{2}^2\angbb{3}{4}^2(k_1 - k_2)(k_3 - k_4)\right) + \mathrm{Cyc}[234],
\label{AllplusQ}
}
matching the general structure in equation (\ref{eqn:JacobiFailure}). The numerators satisfying kinematic Jacobi relations are then obtained by plugging \eqref{eqn:AllPlusNumerator} (and the analogous formulas for $n_t$ and $n_u$) and \eqref{AllplusQ} into \eqref{ntilde}.

Finally, plugging the kinematic numerators into the first line of \eqref{eq:ymampads} gives a very compact formula for the all-plus AdS$_4$ amplitude:
\eqs{
\ang{++++} &= \frac{1}{8k_1k_2k_3k_4}\frac{1}{s}\angbb{1}{2}\angbb{3}{4}\Big[i(\angbb{1}{2}\angbb{4}{1}\angbb{1}{3} + \angb{2}{1}\angbb{3}{2}\angbb{2}{4})\\
&\qquad - \modks \left(\angbb{2}{3}\angbb{4}{1} - \angbb{1}{3}\angbb{2}{4}\right) - \frac{1}{\modks}\angbb{1}{2}\angbb{3}{4}(k_1 - k_2)(k_3 - k_4)\Big] + 2\leftrightarrow4.
\label{eqn:AllPlusCorrelator}
}
Using \eqref{flatspacelimit}, we see that it manifestly vanishes in the flat space limit, as expected.

\subsection{N$^{-1}$MHV}
Next, let us consider the $-++\,+$ case. Converting equation (\ref{eqn:sNumeratorAll}) to spinor notation for the case where particle 1 has negative helicity and the rest positive gives
\eqs{
n_s^{-+++} =& \frac{1}{8k_1k_2k_3k_4}\frac{1}{E} \Big [\angb{1}{2}^2 \angbb{3}{4}^2
(\ang{13}\angbb{1}{3} + \ang{24}\angbb{2}{4}  -\frac{E}{\modks}(k_1-k_2)(k_3-k_4)-E \modks )  \\
&+2s \angb{1}{3} \angbb{2}{4} \angb{1}{2} \angbb{3}{4} \Big ]+s W_s^{B,-+++} ,
}
where
\eqs{
W_s^{B,-+++}=\frac{i}{8k_1k_2k_3k_4E}\frac{2(E - 2k_1)}{s}\angb{1}{2}\angbb{3}{4}\big(\ang{12}\angbb{2}{3}\angbb{2}{4} + \angbb{2}{1}\angb{1}{4}\angb{1}{3}\big),
}
as shown in Appendix \ref{app:BTerms}. As in the all-plus case, we use 4-point spinor and Schouten identities to combine the terms with a $1/E$ pole. In the end, the kinematic numerator can be written as
\eqs{
n_s^{-+++} = &= \frac{1}{8k_1k_2k_3k_4}\angb{1}{2}\angbb{3}{4}\Big[\frac{4ik_1}{E}\angbb{1}{2}\angb{1}{4}\angb{1}{3}+ i\left(\ang{12}\angbb{2}{3}\angbb{2}{4} + \angbb{2}{1}\angb{1}{4}\angb{1}{3}\right)\\
&\qquad +  (\modks+2k_1) \left(\angb{1}{4}\angbb{2}{3} - \angb{1}{3}\angbb{4}{2}\right) - \frac{1}{\modks} \angb{1}{2}\angbb{3}{4}(k_1 - k_2)(k_3 - k_4)\Big].
\label{nsmppp}
}
Using this result and \eqref{eqn:numeratorexchange}, we find that the sum over kinematic numerators is
\eqs{
Q^{-+++} = \frac{1}{8k_1k_2k_3k_4}\left(\modks \left(\angb{1}{3}^2\angbb{2}{4}^2 - \angb{1}{4}^2\angbb{2}{3}^2 \right) - \frac{1}{\modks}\angb{1}{2}^2\angbb{3}{4}^2(k_1 - k_2)(k_3 - k_4)\right) + \mathrm{Cyc}[234].
}
Plugging the above equation along with \eqref{nsmppp} (and the analogous numerators in the $t$- and $u$-channels) into \eqref{ntilde} then gives numerators satisfying kinematic Jacobi relations.

Unlike the all-plus case, the kinematic numerators do not vanish in the flat-space limit, although this will hold for the full AdS$_4$ amplitude. The flat space numerators in this case can be derived from the self-dual sector of Yang-Mills theory \cite{Monteiro:2011pc}. Plugging the kinematic numerators into the first line of \eqref{eq:ymampads}, the amplitude can be reduced to the following concise expression:
\eqs{
\ang{-++\,+} &= \frac{1}{8k_1k_2k_3k_4}\frac{1}{E s}\angb{1}{2}\angbb{3}{4}\Big[4ik_1\angbb{1}{2}\angb{1}{4}\angb{1}{3}\\
&\qquad + i E\left(\ang{12}\angbb{2}{3}\angbb{2}{4} + \angbb{2}{1}\angb{1}{4}\angb{1}{3}\right)\\
&\qquad + E (\modks+2k_1) \left(\angb{1}{4}\angbb{2}{3} - \angb{1}{3}\angbb{4}{2}\right)\\
&\qquad - \frac{1}{\modks}E \angb{1}{2}\angbb{3}{4}(k_1 - k_2)(k_3 - k_4)\Big] + 2\leftrightarrow4.
\label{eqn:SingleMinusResult}
}
To see that the pole in $E$ is in fact spurious, we combine the two terms over a single denominator. Collecting the terms that go as $k_1/E$, we get a numerator proportional to
\eqs{
-4k_1 \angb{1}{2}\angb{1}{3}\angb{1}{4}&\,\big(\angbb{1}{2}\angbb{3}{4}t + \angbb{2}{3}\angbb{4}{1}s\big)\\
\qquad &= -2k_1 \angb{1}{2}\angb{1}{3}\angb{1}{4}E\Big(E\left(\angbb{1}{2}\angbb{3}{4} + \angbb{2}{3}\angbb{4}{1}\right)\\
&\qquad +2(\modkt\angbb{1}{2}\angbb{3}{4} + \modks\angbb{2}{3}\angbb{4}{1}) + i(\angb{3}{1}\angbb{2}{3}\angbb{3}{4} + \angb{1}{3}\angbb{4}{1}\angbb{1}{2})\Big),
}
so there is no pole in $E$ and the flat space limit vanishes. We can write the amplitude in a way that manifestly has the correct flat space limit as follows:
\eqs{
\ang{-++\,+} &= \frac{1}{8k_1k_2k_3k_4}\frac{1}{s}\angb{1}{2}\angbb{3}{4}\Big[i (\ang{12}\angbb{2}{4}\angbb{2}{3} + \angbb{2}{1}\angb{1}{3}\angb{1}{4})\\
&\qquad +  (\modks+2k_1) (\angb{1}{4}\angbb{2}{3} - \angb{1}{3}\angbb{4}{2})\\
&\qquad-\frac{1}{\modks} \angb{1}{2}\angbb{3}{4}(k_1 - k_2)(k_3 - k_4)\\
&\qquad+\frac{2i k_1 \angb{1}{3}\angb{1}{4}}{t}\left(\angbb{1}{2}(E + 2\modkt) + i \angb{4}{2}\angbb{4}{1}\right)\Big] + 2\leftrightarrow4,
}
where we have used Schouten identities and collected the terms proportional to $\angb{1}{2}\angbb{3}{4}$ to identify a symmetry under $2\leftrightarrow4$.

\subsection{MHV}
Let us now consider the $-+-\,+$ case. By substituting spinors into equation \eqref{eqn:sNumeratorAll}, we get (after some manipulations and identities)
\eqs{
n_s^{-+-+} &= \frac{1}{8k_1k_2k_3k_4}\Big[\frac{4}{E}(k_1k_4 + k_2k_3)\angb{1}{2}\angb{3}{4}\ang{13}\angbb{2}{4}\\
&\qquad + i \frac{(E - 2k_1 - 2k_3)}{E}\angb{1}{2}\angb{3}{4}(\ang{21}\angbb{2}{4}\angb{3}{2} + \angbb{1}{2}\ang{31}\angb{1}{4})\\
&\qquad + \modks \left(\ang{13}^2\angbb{4}{2} - \angb{1}{4}^2\angb{3}{2}^2\right)\\
&\qquad-\frac{1}{\modks}\angb{1}{2}^2\angb{3}{4}^2(k_1 - k_2)(k_3 - k_4)\Big],\\
\label{eqn:MHVnumeratorS}
n_t^{-+-+} &= -n_s^{-+-+}\big|_{2\leftrightarrow4}.
}
The final two lines of this can be seen directly from equation \eqref{eqn:sNumeratorAll} after substituting for polarisation vectors and using equation \eqref{eqn:genMandelS} to combine over a common denominator. The second line comes from the $W^B_s$ term, specialised to the $-+-+$ case. This is covered in more detail in Appendix \ref{app:BTerms}. This just leaves the first line, which is analogous to the all-plus and single minus cases, but with a few extra Schouten identities needed to rewrite the pole in $E$. The `extra' pieces we see compared to equation \eqref{eqn:AllPlusNumerator} are what will give the non-zero MHV flat space amplitude.

The $u$-channel numerator cannot simply be obtained by applying \eqref{eqn:numeratorexchange} to \eqref{eqn:MHVnumeratorS} as we need to exchange particles of different helicity. Instead, we must first apply \eqref{eqn:numeratorexchange} to \eqref{eqn:sNumeratorAll} in terms of general polarisation vectors, and then convert the resulting expression for $n_u$ to spinor notation using methods similar to those used to obtain \eqref{eqn:MHVnumeratorS}. In the end, we find
\eqs{
n_u^{-+-+} &= \frac{1}{8k_1k_2k_3k_4}\ang{13}\angbb{4}{2}\Big[\frac{4}{E}\big((k_1k_2 + k_3k_4)\angb{1}{4}\angb{3}{2} + (k_1k_4 + k_2k_3)\angb{3}{4}\angb{1}{2}\big)\\
&\qquad + i \frac{(E - 2k_1 - 2k_3)}{E}\left(\angb{1}{3}\angb{3}{2}\angb{3}{4} + \angb{3}{1}\angb{1}{4}\angb{1}{2}\right)\\
&\qquad + \modku \left(\angb{1}{4}\angb{3}{2} + \angb{1}{2}\angb{3}{4}\right )\\
&\qquad-\frac{1}{\modku}\ang{13}\angbb{4}{2}(k_1 - k_3)(k_4 - k_2)\Big].
\label{eqn:MHVnumeratorU}
}
Adding up the kinematic numerators or directly converting \eqref{eqn:JacobiFailure} to spinor notation, we find that
\eqs{
Q^{-+-+} &= \frac{1}{8k_1k_2k_3k_4}\left(\modks \left(\ang{13}^2\angbb{4}{2}^2 - \angb{1}{4}^2\angb{3}{2}^2 \right) - \frac{1}{\modks}\angb{1}{2}^2\angb{3}{4}^2(k_1 - k_2)(k_3 - k_4)\right) + \mathrm{Cyc}[234],
\label{eqn:MHVQ}
}
where the permutations act on the bars as well as the particle labels. Plugging \eqref{eqn:MHVnumeratorS}-\eqref{eqn:MHVQ} into \eqref{ntilde} then gives the numerators which satisfy the kinematic Jacobi relation.

Moreover, plugging the above kinematic numerators into the first line of \eqref{eq:ymamp}, we obtain the following remarkably compact formula for the MHV amplitude in AdS$_4$:
\eqs{
\ang{-+-+} &=\frac{n_s^{-+-+}}{s} - \frac{n_t^{-+-+}}{t}\\
&= \frac{1}{8k_1k_2k_3k_4}\frac{1}{Es}\angb{1}{2}\angb{3}{4}\Big[4(k_1k_4 + k_2k_3)\ang{13}\angbb{2}{4}\\
&\qquad\qquad + i (E - 2k_2 - 2k_4)(\ang{12}\angbb{2}{4}\angb{3}{2} + \angbb{2}{1}\ang{31}\angb{1}{4})\\
&\qquad\qquad + E \modks (\ang{13}\angbb{2}{4} - \angb{1}{4}\angb{3}{2})\\
&\qquad\qquad-\frac{1}{\modks}E \angb{1}{2}\angb{3}{4}(k_1 - k_2)(k_3 - k_4)\Big] + 2\leftrightarrow4.
\label{eqn:AltMHVResult}
}
This formula represents substantial progress beyond the formula first obtained in the pioneering work \cite{Raju:2012zs} which contained many more terms and several functions of spinor brackets in the denominators, whereas the denominators of \eqref{eqn:AltMHVResult} contain only $k_i$ and $k_{\underline{ij}}$. The key to obtaining such a simple expression was to focus on simplifying the kinematic numerators and make use of numerous spinor identities derived in Appendix \ref{app:spinors}. It would be interesting to see if this simplicity extends to higher-point amplitudes, and in particular if there is some analogue of the Parke-Taylor formula  \cite{Parke:1986gb} for MHV amplitudes in AdS$_4$.

As in the single-minus case, this does not have the flat space limit manifest until we combine parts from the two numerators. If consider only the terms with a pole in $E$, we get
\eqs{
\lim_{E\rightarrow0}\ang{-+-+} &=\lim_{E\rightarrow0} \frac{1}{8k_1k_2k_3k_4}\frac{1}{Est}\Big[\\
	&\qquad 4\ang{13}\angbb{2}{4}\big(t \angb{1}{2}\angb{3}{4}(k_1k_4 + k_2k_3)\\
		&\qquad\qquad -s\angb{1}{4}\angb{3}{2}(k_1k_2 + k_3k_4)\big)\\
	&\qquad+2(k_2 + k_4)\big(t\angb{1}{2}\angb{3}{4}(\angbb{1}{2}\ang{13}\angb{1}{4} + \ang{12}\angb{3}{2}\angbb{2}{4})\\
		&\qquad\qquad+s\angb{1}{4}\angb{3}{2}(\angbb{1}{4}\ang{13}\angb{1}{2} + \ang{14}\angb{3}{4}\angbb{4}{2})\big)\Big],\\
&=-\frac{2}{E}\frac{\ang{13}^2\angbb{2}{4}^2}{\ang{12}\angbb{1}{2}\ang{23}\angbb{2}{3}},
}
where we recognise the final line as the familiar four point flat space amplitude. The intermediate steps involve many spinor and Schouten identities but the general idea is to start with the lowest order terms in $k_i$ and apply identities that generate higher powers (avoiding any $k_i^2$) along with corrections of order $E$, which are discarded. Eventually the only remaining contribution is a numerator proportional to $k_1k_2k_3k_4$ (which cancels the factors from the polarisation vectors) and corrections subleading in $E$, which we have omitted here.

Amplitudes with a $u$-channel contribution are less compact. For example we can obtain the $\ang{--++}$ amplitude from
\eq{
\ang{--++} = \frac{n_s^{--++}}{s} - \frac{n_t^{--++}}{t},}
\eq{\mathrm{where}\quad n_s^{--++}= -n_u^{-+-+}\Big|_{2\leftrightarrow 3},\quad n_t^{--++} = -n_t^{-+-+}\Big|_{2\leftrightarrow 3},
}
where we exchange particle label but not helicity labels (i.e. bars on spinors), and the numerators are as defined in \eqref{eqn:MHVnumeratorS} and \eqref{eqn:MHVnumeratorU}. This operation should be thought of as a relabelling and can easily be seen by taking the diagrams in Figure \ref{fig:numerators}, dressing with helicities and applying the relabellings. For completeness, we also note that
\eq{
n_u^{--++} = -n_s^{-+-+}\Big|_{2\leftrightarrow 3}.
}
We can then see that the sum over kinematic numerators for $--++$ is given by
\eq{
Q^{--++} = -Q^{-+-+}\Big|_{2\leftrightarrow3}.
}

\section{Relation to 3d Conformal Correlators}
\label{sec:Reconstruction}

The AdS amplitudes we computed in previous sections are closely related to 3d CFT correlators in momentum space. In particular, they encode the transverse parts of the correlators from which the full correlators can be reconstructed using Ward identities, as we will explain in this section. The transverse part of a correlator can be recovered from an AdS amplitude by stripping off the polarisation vectors and replacing them with projection tensors:
\eqs{
\ang{j_1 \dots j_n}_\pi &= \ang{j_1 \dots j_n}\big|_{\epsilon_i \to \pi_i},\\
\mathrm{where},\quad \pi_i^{jk} &= \eta ^{jk}-\frac{k_i^{j}k_i^{k}}{k_i^2}.
}
In the above formula for transverse projection tensors, upper indices are 3d Lorentz indices while lower indices are particle labels.

The full correlator, which we will denote as $\ang{J_1 \dots J_n}$, can then be reconstructed using the transverse Ward identity. For spin-1 currents with color indices, this is well known \footnote{This Ward identity can be found in many text books. For a discussion in the context of cosmology see \cite{Maldacena:2011nz,Baumann:2020dch}.}:
\eqs{
k_{1i_1}\ang{J^{a_1i_1}(k_1) J^{a_2i_2}(k_2) \dots J^{a_ni_n}(k_n)} &= -if^{a_1a_2b}\ang{J^{bi_2}(k_2 + k_1) \dots J^{a_ni_n}(k_n)} \\
& \cdots\, -i f^{a_1a_nb}\ang{J^{a_2i_2}(k_2) \dots J^{bi_n}(k_n + k_1)},
}
where we have defined the structure constants such that $[T^a,T^b] = -if^{abc}T^c$ and $\mathrm{tr}\left(T^aT^b\right) = \delta^{ab}$. The right hand side of this equation corresponds to the standard sum over contact terms in position space. In the flat space limit, the contact terms don't contribute because they are constructed from lower-point correlators and therefore do not contain a pole in $E=\sum_{i=1}^{n} k_i$. Hence, the Ward identity for conformal correlators reduces to a Ward identity for flat space scattering amplitudes in one higher dimension. For color-ordered correlators, there are only two contact terms:
\eqs{
k_{1i_{1}}\ang{J^{i_{1}}(k_{1})J^{i_{2}}(k_{2})\dots J^{i_{n}}(k_{n})}=\left(-1\right)^{n}\left(\left\langle J^{i_{2}}(k_{2})\dots J^{i_{n}}(k_{n}+k_{1})\right\rangle -\left\langle J^{i_{2}}(k_{2}+k_{1})\dots J^{i_{n}}(k_{n})\right\rangle \right).
}
The full correlator can then be obtained from the transverse part by adding terms proportional to lower-point correlators which give the required contact terms when contracted with the momentum $k_1^i$.

The reconstruction of 3-point correlators was spelled out in \cite{Bzowski:2013sza}, and we will review it here for completeness. We will then present a new formula for 4-point correlators, which is the main focus of this paper. At three points the Ward identity can be written as
\begin{align}
    k_{1i} \ang{ J^i (\vec{k}_1) J^j (\vec{k} _2) J^k (\vec{k} _3)} =\ang{ J^j(-\vec{k}_3) J^k(\vec {k}_3)} -\ang{ J^j(\vec{k}_2) J^k (-\vec{k}_2)},
\label{eqn:3ptWI}
\end{align}
where the 2-point function is fixed by conformal Ward identities to be \cite{Bzowski:2013sza}:
\eq{
    \ang{ J^{i} (\vec{k}) J^{j} (-\vec{k})} =c_J k\, \pi ^{ij},
}
where $c_J$ is a normalisation. The full correlator can be recovered by adding terms to the transverse piece such that \eqref{eqn:3ptWI} is satisfied, symmetrising (recalling that color-ordered correlators are cyclically symmetric), and adding further terms to cancel the new terms which arise after symmetrising and contracting with $k_1$. In the end this gives
\eqs{
    &\la J^{i}(\vec{k}_1)J^{j}(\vec{k}_2)J^{k}(\vec{k}_3)\ra= \ang{ j^{i }(\vec{k}_1)j^{j}(\vec{k}_2)j^{k}(\vec {k}_3)}_\pi \\
    & + \left[ \frac{k_1^{i}}{k_1^2} \left( \la J^{j}(-\vec{k}_3) J^{k}(\vec{k}_3)\ra - \la J^{j}(\vec{k}_2) J^{k}(-\vec{k}_2)\ra\right)+ \frac{k_1^{i}k_2^{j}}{k_1^2k_2^2}\,k_{1a} \ang{ J^{a} (\vec{k}_3)J^{k}(-\vec{k}_3) }\right] + \mathrm{Cyc[123]},
\label{Full3pt}
}
where $\ang{jjj}_\pi$ is obtained from the 3-point AdS amplitude in \eqref{eqn:3ptAdSAmp}. The cyclic sum includes permutations of $i,j,k$ along with the particle labels.

At higher points, the procedure for reconstructing the full correlator from the transverse piece by adding terms to solve the Ward identity, symmetrising, and adding new terms to cancel unwanted contributions must be iterated. At four-points, this will involve adding numerous terms with both 2-point and 3-point correlators. Using \eqref{eqn:3ptWI} to tidy up, we are left with
\eqs{
\ang{J^i_1J^j_2J^k_3J^l_4} &= \ang{j^i_1j^j_2j^k_3j^l_4}_\pi + \bigg[\frac{k_1^i}{k_1^2}\left(\ang{J^j(k_2)J^k(k_3)J^l(k_4 + k_1)}-\ang{J^j(k_1+k_2)J^k(k_3)J^l(k_4)} \right)\\
&\qquad+\frac{k^i_1k^j_2}{k_1^2k_2^2}\Big(\ang{J^k(\vec{k}_3)J^l(\vec{-k}_3)} - \ang{J^k(\vec{k}_2 + \vec{k}_3)J^l(-\vec{k}_2 - \vec{k}_3)} \\
&\qquad\qquad\qquad\qquad + k_{2a}\ang{J^a(k_1+k_2)J^k(k_3)J^l(k_4)}\Big)\\
&\qquad +\frac{k^i_1k^k_3}{2k_1^2k_3^2}\Big(\ang{J^j(\vec{k}_2 + \vec{k}_3)J^l(-\vec{k}_2 - \vec{k}_3)} + \ang{J^j(\vec{k}_3 + \vec{k}_4)J^l(-\vec{k}_3 - \vec{k}_4)}\\&\qquad\qquad\qquad\qquad - \ang{J^j(\vec{k}_2)J^l(\vec{-k}_2)} - \ang{J^j(\vec{k}_4)J^l(\vec{-k}_4)}\Big)\\
&\qquad + \frac{k_2^jk_3^kk_4^l}{k_2^2k_3^3k_4^4}k_{3a}\Big(\ang{J^a(\vec{k}_1)J^i(\vec{-k}_1)} - \ang{J^a(\vec{k}_2 + \vec{k}_3)J^i(-\vec{k}_2 - \vec{k}_3)}  \\
&\qquad\qquad\qquad\qquad - \ang{J^a(\vec{k}_3 + \vec{k}_4)J^i(-\vec{k}_3 - \vec{k}_4)}\Big)\\
&\qquad + \frac{k^i_1k^j_2j^k_3k^l_4}{4k_1^2k_2^2k_3^2k_4^2}k_{1a}k_{3b}\left(\ang{J^a(\vec{k}_2 + \vec{k}_3)J^b(-\vec{k}_2 - \vec{k}_3)} + \ang{J^a(\vec{k}_3 + \vec{k}_4)J^b(-\vec{k}_3 - \vec{k}_4)}\right)\bigg]\\
& \qquad\qquad \qquad\qquad + \mathrm{Cyc}[1234],
}
where the first term is obtained from the 4-point AdS amplitudes computed in the previous sections.

\section{Conclusion} \label{conclusion}

In this paper we explored how CK duality is realised for tree-level 4-point YM amplitudes in AdS$_4$. In particular, we found a decomposition of these amplitudes into kinematic numerators analogous to those of flat space amplitudes. In contrast to flat space, we find that numerators in AdS$_4$ do not generically satisfy kinematic Jacobi identities. We also find that color-ordered amplitudes obey a deformed BCJ relation which reduces to the usual one in the flat space limit (see \eqref{bcjads}). On the other hand, the numerators can be shifted in such a way that the amplitudes are preserved. Using this generalised gauge symmetry, there is a unique choice of numerators which do obey the kinematic Jacobi identity, given in \eqref{ntilde}.  We then recast these results in spinor notation to obtain concise new formulae for all helicity configurations using numerous spinor identities derived in Appendix \ref{app:spinors}, and explain how to reconstruct 3d conformal correlators of conserved currents from AdS$_4$ amplitudes using Ward identities.

The numerators which obey kinematic Jacobi relations may play an important role in obtaining gravitational amplitudes in AdS$_4$ via a double copy procedure. In more detail, we can construct an object analogous to the flat space gravitational amplitude in \eqref{gravdouble} by squaring these numerators:
\eqs{
\ang{t_1t_2t_3t_4} &=\frac{k_1 k_2 k_3 k_4}{E} \left(\frac{\tilde{n}_s^2}{s} + \frac{\tilde{n}_t^2}{t} + \frac{\tilde{n}_u^2}{u} \right),\\
&=\frac{k_1 k_2 k_3 k_4}{E} \left(\frac{n_s^2}{s} + \frac{n_t^2}{t} + \frac{n_u^2}{u} - \frac{1}{\xi}Q^2 \right),
\label{adsdouble}
}
where $s$, $t$ and $u$ are generalised Mandelstam variables as given in \eqref{eqn:GenMandelstam}. We label this object $\ang{t_1t_2t_3t_4}$ to denote the transverse traceless part of a stress tensor correlator, dual to a gravitational amplitude in the bulk. From this quantity, it should be possible to reconstruct the full 4-point correlator in momentum space using Ward identities \cite{Dymarsky:2013wla,Bzowski:2017poo}. We have also included a prefactor with an additional pole in $E$ since in the flat space limit, 3d stress tensor correlators behave as follows \cite{Raju:2012zr}:
\eq{
\lim_{E\rightarrow0}\left\langle t_{1}t_{2}t_{3}t_{4}\right\rangle =\frac{k_{1}k_{2}k_{3}k_{4}}{E^{3}}\mathcal{M}_{4},
}
where $\mathcal{M}_4$ is the 4d graviton amplitude in flat space. Using the kinematic Jacobi relation in \eqref{adskinematicjacobi}, the deformed BCJ relation in \eqref{deformedbcj2}, and the relation among generalised Mandelstam variables in \eqref{stuxi}, we find that \eqref{adsdouble} implies a deformed KLT relation between the stress tensor correlator and a product of current correlators:
\eq{
\ang{t_1t_2t_3t_4} =-\frac{k_1 k_2 k_3 k_4}{E} \left(s\ang{j_1j_2j_3j_4}\ang{j_1j_2j_4j_3} + \xi\frac{\tilde{n}_t\tilde{n}_u}{tu}\right).
\label{kltdef}
}
In the flat space limit, this reduces to the standard KLT relation in \eqref{kltrelat}. It would be interesting to explicitly compute a 4-point graviton amplitude in AdS$_4$ and see how it compares to \eqref{adsdouble} or \eqref{kltdef}. Note that any corrections to these equations must be subleading in the flat space limit, so the main question is to understand the structure of such corrections.

In addition to comparing \eqref{adsdouble} to explicit Witten diagram calculations, there are several other directions for future work. Perhaps the most obvious is to see if the CK duality in AdS$_4$ can be extended to higher points. At first sight, the 5-point expressions obtained in \cite{Albayrak:2018tam} look somewhat formidable, but it may be possible to simplify them by converting to spinor notation and generalising the spinor identities in Appendix \ref{app:spinors} to higher points. As we noted in section \ref{overview}, the CK duality in AdS has some similarities to that of massive amplitudes in flat space. On the other hand, it has recently been shown that the massive double copy generally introduces unphysical singularities above 4-points \cite{Johnson:2020pny}, so it would also be interesting to see if this can be avoided in AdS. In flat space, many aspects of the double copy become manifest by expressing amplitudes in terms of scattering equations \cite{Cachazo:2013hca,Mason:2013sva}. Recently, analogous equations were proposed in AdS \cite{Roehrig:2020kck,Eberhardt:2020ewh}, so it would be interesting to see how that approach is related to the one developed in this paper. This may also suggest how to formulate the KLT relations for string theory in AdS.

It would also be interesting to explore to what extent the CK duality and double copy hold for generic theories in AdS$_4$, or equivalently correlators in generic 3d CFT's. Perhaps the best way to approach this question would be to find general solutions to the conformal Ward identities in momentum space for $n$-point correlators of currents and stress tensors analogous to the general solution for scalar correlators recently obtained in terms of Feynman integrals in \cite{Bzowski:2019kwd}. It may then be possible to look for double copy structure by studying their leading singularities, as was shown at 3-points in \cite{Lipstein:2019mpu}. Finally, it would be very interesting to adapt this story to dS$_4$ with the goal of seeing new mathematical structure in cosmological observables. Some work on the double copy in dS$_4$ has been carried out in \cite{Li:2018wkt,Fazio:2019iit}.

\begin{center}
\textbf{Acknowledgements}
\end{center}
We thank Henrik Johansson and Paul McFadden for many useful comments on the manuscript. AL is supported by a Royal Society University Research Fellowship. CA's studentship is also funded by the Royal Society.

\appendix
\section{Witten Diagrams} \label{app:wittdiag}
In this section we will review 4-point color-ordered Witten diagrams for YM in AdS$_4$. We follow the approach developed in \cite{Liu:1998ty,Raju:2011mp,Albayrak:2018tam}, and will review some basic formulas to make the discussion self-contained.

We will work with the following metric:
\eq{
ds^{2}=\frac{\eta_{ij}dx^{i}dx^{j}+dz^{2}}{z^{2}},
}
where the boundary metric $\eta_{ij}={\rm diag\left\{ -1,+1,+1\right\} }$. In axial gauge, the bulk-to-bulk propagator in momentum space is given by
\begin{equation}
    \mathcal{G} _{ij} (z,z',\vec{k})=-i \int_{0}^{\infty} p  dp \frac{z^{\frac{1}{2}}\boldsymbol{J}_{\frac{1}{2}}(pz)\boldsymbol{J}_{\frac{1}{2}}(pz')(z')^{\frac{1}{2}}}{\boldsymbol{k}^2+p^2} \left(\eta _{ij}+\frac{\boldsymbol{k}_i\boldsymbol{k}_j}{p^2}\right),
\end{equation}
where $\vec{k}$ is the momentum flowing through the propagator along the boundary directions, and $\boldsymbol{J}_{\nu}$ is a Bessel function of the first kind. For space-like momenta along the boundary, the bulk-to-boundary propagator is given by
\begin{equation}
    A_m(z,\vec{k})=\epsilon _m \sqrt{\frac{2 k}{\pi}} z^{\frac{1}{2}} \boldsymbol{K}_{\frac{1}{2}}(kz),
\end{equation}
where the polarisation $\epsilon$ does not have a radial component so $m$ is a 3d Lorentz index, and $\boldsymbol{K}_{\nu}$ is a modified Bessel function of the second kind.

The vertices have the same structure as those in flat space but the indices only run over the boundary directions since we are in axial gauge:
\eqs{
    V_{jkl}(\vec{k}_1,\vec{k}_2,\vec{k}_3) &= \frac{i}{\sqrt{2}}\left(\eta _{jk}(\vec{k}_1-\vec{k}_2)_l +\eta _{kl}(\vec{k}_2-\vec{k}_3)_j+\eta _{lj}(\vec{k}_3-\vec{k}_1)_k\right),\\
    V^{jklm}&=i \eta ^{jl}\eta ^{km}-\frac{i}{2} \left(\eta ^{jk}\eta ^{lm}+ \eta ^{jm}\eta ^{kl}\right),
\label{ymvertices}
}
where we have set the YM coupling $g=1$. When computing Witten diagrams, we must integrate over the radial coordinates $z$ of each interaction vertex. In doing so, a factor of $z^{-4}$ coming from $\sqrt{-g}$ will be cancelled by $z^4$ coming from two inverse metrics used to contract the Lorentz indices in each interaction vertex.

Using the rules given above, we can write down the three-point AdS$_4$ amplitude as follows:
\begin{align}
\ang{jjj}&=V_{123}(\boldsymbol{k}_1,\boldsymbol{k}_2,\boldsymbol{k}_3)\int _0^{\infty} dz \left( \sqrt{\frac{2k_1}{\pi}}z^{\frac{1}{2}}\boldsymbol{K}_{\frac{1}{2}}(k_1z)\right ) \left( \sqrt{\frac{2k_2}{\pi}}z^{\frac{1}{2}}\boldsymbol{K}_{\frac{1}{2}}(k_2z)\right )  \left( \sqrt{\frac{2k_3}{\pi}}z^{\frac{1}{2}}\boldsymbol{K}_{\frac{1}{2}}(k_3z)\right ),   \nonumber \\
    &= \frac{1}{\sqrt{2} E}\left(\epsilon_1\!\cdot\! \epsilon_2\, (\vec{k}_1-\vec{k}_2)\!\cdot\!\epsilon_3 + \mathrm{Cyc}[123]\right).
\label{eqn:3ptAdSAmp}
\end{align}
where $V_{123}= \epsilon_1^i \epsilon_2^i \epsilon_3^i V_{ijk}$.

The color-ordered four-point AdS amplitude comes from three Witten diagrams, as shown in Figure \ref{fig:WittenDiag4ptAmp}. In particular, there are $s$- and $t$-channel exchanges, and a contact diagram:
\eq{
\ang{jjjj}=W_s + W_t + W_c.
}
The quantity given by multiplying vertices and propagators together and summing diagrams is usually denoted as $i\mathcal{A}$. Since our final expressions will contain overall factors of $i$, we drop these on both sides. The $s$-channel diagram is then given by
\eq{
    W_s=\int p dp\, dz dz'\, \mathrm{K}\mathrm{K}\mathrm{J}(k_1,k_2,p,z)\frac{M^{1234}(\boldsymbol{k}_1,\vec{k}_2,\vec{k}_3,\vec{k}_4)}{(\vec{k}_{12}^2+p^2)}\mathrm{K}\mathrm{K}\mathrm{J}(k_3,k_4,p,z'),
\label{wsintegral}
}
where
\eq{
    M^{ijkl}(\vec{k}_1,\vec{k}_2,\vec{k}_3,\vec{k}_4)=-V^{ijm}(\vec{k}_1,\vec{k}_2,-\vec{k}_{12})\left (\eta _{mn}+\frac{\vec{k}_m\vec{k}_n}{p^2}\right ) V^{kln}(\vec{k}_3,\vec{k}_4,\vec{k}_{12}),
}
and
\eq{
\mathrm{K}\mathrm{K}\mathrm{J}(p,r,s,z) = \sqrt{\frac{4pr}{\pi^2}}z^{3/2}K_{\frac{1}{2}}(pz)K_{\frac{1}{2}}(rz)J_{\frac{1}{2}}(sz).
}
The integrals in \eqref{wsintegral} were performed in \cite{Albayrak:2018tam} with the final result
\eq{
W_s=-\frac{V^{12m}(\boldsymbol{k}_1,\boldsymbol{k}_2,-\boldsymbol{k}_{12})V^{34n}(\boldsymbol{k}_3,\boldsymbol{k}_4,\boldsymbol{k}_{12})}{k_{1234}k_{12\underline{12}}k_{34\underline{12}}} \left (\eta _{mn}+\frac{k_{1234\underline{12}}(\boldsymbol{k} _{12})_m(\boldsymbol{k} _{12})_n}{k_{12}k_{34}k_{\underline{12}}}  \right),
\label{wsdiag}
}
where $V^{12m}=\epsilon _1^i \epsilon _2^j V_{ijm}$. We can simplify \eqref{wsdiag} by applying momentum conservation and other identities at each vertex:
\begin{align}
    V^{12m} \cdot (\vec{k}_{12})_m&=\frac{i}{\sqrt{2}}(\epsilon _1 \!\cdot\! \epsilon _2\, (k_1^2-k_2^2)+2 \vec{k} _2 \!\cdot\! \epsilon _1\, \epsilon _2\!\cdot\! \vec{k}_{12} - 2\vec{k} _1 \!\cdot\! \epsilon _2\, \epsilon _1 \!\cdot\! \vec{k} _{12}) \nonumber \\
    &=\frac{i}{\sqrt{2}}\epsilon _1 \!\cdot\! \epsilon _2\, (k_1^2-k_2^2),
\end{align}
since $\vec{k} _i \!\cdot\! \epsilon _i=0$. Moreover, using momentum conservation we get
\begin{align}
    V^{12m} \cdot V^{34}_m=&-\frac{1}{2}(\epsilon _1 \cdot \epsilon _2 \epsilon _3\cdot \epsilon _4)(\vec{k}_1-\vec{k}_2)\!\cdot\!  (\vec{k}_3-\boldsymbol{k}_4) \nonumber \\
    &-\epsilon _1\!\cdot\!  \epsilon _2\, \vec{k} _4 \!\cdot\!  \epsilon _3 (\vec{k}_1-\vec{k}_2)\!\cdot\!  \epsilon _4+\epsilon _1\!\cdot\!  \epsilon _2\, \vec{k} _3 \!\cdot\!  \epsilon _4 (\vec{k}_1-\vec{k}_2)\!\cdot\!  \epsilon _3 \nonumber\\
    &-\epsilon _3\!\cdot\!  \epsilon _4\, \vec{k} _2 \!\cdot\!  \epsilon _1 (\vec{k}_3-\vec{k}_4)\!\cdot\!  \epsilon _2+\epsilon _3\!\cdot\!  \epsilon _4\, \vec{k} _1 \!\cdot\!  \epsilon _2 (\vec{k}_3-\vec{k}_4)\!\cdot\!  \epsilon _1 \nonumber\\
    &-2\epsilon _2\!\cdot\!  \epsilon _4\, \vec{k} _2 \!\cdot\!  \epsilon _1\, \vec{k}_4\!\cdot\!  \epsilon _3+2\epsilon _2\!\cdot\!  \epsilon _3\, \vec{k} _2 \!\cdot\!  \epsilon _1\, \vec{k}_3\!\cdot\!  \epsilon _4 \nonumber \\
    &+2\epsilon _1\!\cdot\!  \epsilon _4\, \vec{k} _1 \!\cdot\!  \epsilon _2\, \vec{k}_4\!\cdot\!  \epsilon _3 -2\epsilon _1\!\cdot\!  \epsilon _3\, \vec{k} _1 \!\cdot\!  \epsilon _2\, \vec{k}_3\!\cdot\!  \epsilon _4.
\end{align}
The terms containing $\vec{k} _i \!\cdot\! \epsilon _j$ will be collected into a single term $W^B_s$, which is given in \eqref{eqn:sChannelBTerm}. In section \ref{app:BTerms}, we will show that this term can be greatly simplified after converting to spinor notation.
The $t$-channel diagram can be obtained by taking the $s$-channel diagram and swapping $2 \leftrightarrow 4$:
\eq{
W_{t}=\left.W_{s}\right|_{2\leftrightarrow4}.
}

Finally, the contact diagram is given by
\eq{
W_c =-i V^{1234}\int_{0}^{\infty} dz \mathrm{K} \mathrm{K} \mathrm{K} \mathrm{K} (k_1,k_2,k_3,k_4,z),
}
where $ V^{1234}=\epsilon_1^i \epsilon_2^j \epsilon_3^k \epsilon_4^l V_{ijkl}$ and
\eq{
\mathrm{K} \mathrm{K} \mathrm{K} \mathrm{K}(p,r,s,t,z) = \sqrt{\frac{16prst}{\pi^4}}z^2 K_{\frac{1}{2}}(pz) K_{\frac{1}{2}}(rz) K_{\frac{1}{2}}(sz) K_{\frac{1}{2}}(tz).
}
After integrating over $z$, one gets
\eq{
W_c=\frac{1}{k_{1234}}\left(\epsilon _1\!\cdot\! \epsilon _3\, \epsilon _2\!\cdot\! \epsilon _4 -\frac{1}{2}(\epsilon _1\!\cdot\! \epsilon _2\, \epsilon _3\!\cdot\! \epsilon _4 +\epsilon _1\!\cdot\! \epsilon _4\, \epsilon _2\!\cdot\! \epsilon _3)\right).
}

\section{Spinor-Helicity Formalism} \label{app:spinors}
In this Appendix, we will review the spinor-helicity formalism for AdS$_4$ and then derive numerous useful identities and describe a general strategy for simplifying Witten diagrams in spinor notation. A null 4-momentum in AdS$_4$ is given by \eqref{eqn:nullMomDefn}. Contracting the 4-momentum with Pauli matrices then gives
\eq{
k^{\alpha\dot{\alpha}} = k^\mu\left(\sigma_\mu\right)^{\alpha\dot{\alpha}} = \begin{pmatrix}
k^0 + ik	&k^1-ik^2\\
k^1+ik^2	&k^0 - ik
\end{pmatrix}.
}
The determinant of this matrix is zero by construction. We can therefore write it in bi-spinor form as shown in \eqref{4dspin} where
\eq{
\lambda^\alpha = \begin{pmatrix}
\sqrt{k^0 + i k}\\
\frac{k^1 - i k^2}{\sqrt{k^0 + i k}}
\end{pmatrix},\qquad
\ltilde{}^{\dot{\alpha}} = \begin{pmatrix}
\sqrt{k^0 + i k}\\
\frac{k^1 + i k^2}{\sqrt{k^0 + i k}}
\end{pmatrix}.
}

Since momentum in the radial direction is not conserved, it is natural to use the unit vector in the radial direction to convert dotted into undotted indices, breaking the 4d Lorentz group to the 3d Lorentz group, as described in section \ref{helicityamp}. Let us therefore define
\eq{
\lbar{i}^\alpha = -\varepsilon^{\alpha\beta}(\sigma^3)_{\beta\dot{\beta}}\ltilde{i}^{\dot{\beta}} \equiv \begin{pmatrix}
-\frac{k^1 + i k^2}{\sqrt{k^0 + i|\mathbf{k}|}}\\
-\sqrt{k^0 + i|\mathbf{k}|}
\end{pmatrix}.
\label{eqn:lambdabar}
}
We are then free to contract all spinors using $\varepsilon_{\alpha\beta}$ according to \eqref{2bracket}. This gives us the special case
\eq{
\angb{m}{m} = \varepsilon_{\alpha\beta}\lambda_m^\alpha\lbar{m}^\beta = -2i k_m.
}
After converting dotted to undotted indices, the 4-momentum can be written in bispinor form according to \eqref{adsbispin}. The inner product of two 4-momenta (taken with respect to a flat metric) is then given by
\eq{
2k_{\mu}  q^\mu = \ang{kq}\ang{\bar{k}\bar{q}}.
}
We can use this to derive the 3d dot product
\eq{
\mathbf{k}\cdot\mathbf{q} = k_{\mu}  q^\mu + k q = \frac{1}{2}\ang{kq}\ang{\bar{k}\bar{q}} - \frac{1}{4}\ang{k\bar{k}}\ang{q\bar{q}}=\frac{1}{4}\left(\left\langle kq\right\rangle \left\langle \bar{k}\bar{q}\right\rangle +\left\langle k\bar{q}\right\rangle \left\langle \bar{k}q\right\rangle \right).
\label{eqn:3dDotProd}
}
The rightmost expression follows directly from the definition of 3-momentum implied by the decomposition in  \eqref{adsbispin}. Its equivalence to the middle expression follows from the Schouten identity in \eqref{shouten}.

Finally, using the definition of polarisation vectors in \eqref{polspin}, we have the following useful formulas for inner products:
\eqs{
2\mathbf{k}_m \cdot\epsilon^+_n = \frac{\ang{m\bar{n}}\ang{\bar{m}\bar{n}}}{ik_n},&\qquad2\mathbf{k}_m \cdot\epsilon^-_n = \frac{\ang{mn}\ang{\bar{m}n}}{ik_j},\\
2\epsilon^+_m\cdot\epsilon^+_n = -\frac{\ang{\bar{m}\bar{n}}^2}{k_mk_n},&\qquad 2\epsilon^-_m\cdot\epsilon^-_n = -\frac{\ang{mn}^2}{k_mk_n},\\
2\epsilon^-_m\cdot\epsilon^+_n &= -\frac{\angb{m}{n}^2}{k_mk_n}.
}

\subsection{Useful Identities}
\label{app:useful identities}
At three points, \eqref{esum} reduces to
\eqs{
\lambda _1^\alpha \bar {\lambda} _1^{\beta}+\lambda _2^\alpha \bar {\lambda} _2^{\beta}+\lambda _3^\alpha \bar {\lambda} _3^{\beta} &=i E \epsilon ^{\alpha \beta},\\
k_1+k_2+k_3&=E,\\
\boldsymbol{k} _1 +\boldsymbol{k} _2 +\boldsymbol{k} _3 &=0.
\label{eqn:dS3ptMomCons}
}
We can derive one identity from the 4-momentum dot product by expanding it in terms of 3-momenta and the radial components, then using momentum conservation
\eqs{
\ang{on}\angbb{o}{n} &= 2k_o^\mu k_{n\mu} = (\mathbf{k}_o + \mathbf{k}_n)^2 + (ik_o + ik_n)^2,\\
&= k_m^2 - (k_n+k_o)^2 = E_p(k_m - k_n - k_o).
}
Other identities can be obtained by contracting the top line of equation (\ref{eqn:dS3ptMomCons}) with various spinors:
\eqs{
\ang{on}\angbb{o}{n} &= E(k_m - k_n - k_o),\\
\ang{mn}\angbb{n}{o} &= iE \angb{m}{o},\\
\ang{mn}\ang{\bar{n}o}& = i\ang{mo}(k_m+k_n-k_o),\\
\angbb{m}{n}\angb{n}{o} &= i\angbb{m}{o}(k_o-k_m-k_n),\\
\angb{m}{n}\angb{n}{o} &= i\angb{m}{o}(k_m + k_o - k_n),
\label{eqn:3ptIdentities}
}
for distinct particle labels $m,n,o$.

We can carry out the same process at four points where there are a few extra possibilities as we have more freedom to contract with different momenta:
\eqs{
\ang{mn}\angbb{m}{n} - \ang{op}\angbb{o}{p} &= E (k_o+k_p-k_m-k_n),\\
\ang{mn}\angbb{n}{p} + \ang{mo}\angbb{o}{p} &= iE \angb{m}{p},\\
\ang{mn}\ang{\bar{n}p} + \ang{mo}\ang{\bar{o}p} &=i\ang{mp}(k_m + k_n + k_o - k_p),\\
\angbb{m}{n}\angb{n}{p} + \angbb{m}{o}\angb{o}{p} &= i\angbb{m}{p}(k_p - k_m - k_n - k_o),\\
\angb{m}{n}\angb{n}{p} + \angb{m}{o}\angb{o}{p} &= i\angb{i}{p}(k_m + k_p - k_n - k_o),\\
\ang{mn}\angbb{m}{n} + \ang{mo}\angbb{m}{o}+\ang{mp}\angbb{m}{p} &= -2 E k_m,
\label{eqn:4ptIdentities}
}
where $m,n,o,p$ are distinct particle labels.

Some of these identities imply additional useful relations. For example, consider the following structures:
\eqs{
A &= \angb{1}{2}\angbb{1}{3}\angbb{4}{1} + \angb{2}{1}\angbb{2}{4}\angbb{3}{2},\\
B &= \angb{3}{4}\angbb{1}{3}\angbb{3}{2} + \angb{4}{3}\angbb{2}{4}\angbb{4}{1}.
}
Using momentum conservation to expand the mixed brackets we then find
\eqs{
A &= \frac{1}{iE}\big(\ang{13}\angbb{3}{2}\angbb{1}{3}\angbb{4}{1} + \ang{14}\angbb{4}{2}\angbb{1}{3}\angbb{4}{1}\\
&\qquad + \ang{23}\angbb{3}{1}\angbb{2}{4}\angbb{3}{2} + \ang{24}\angbb{4}{1}\angbb{2}{4}\angbb{3}{2}\big) = B,
\label{eqn:4ptReflection}
}
where the second equality is obtained by rearranging the terms and applying momentum conservation again. In fact, these types of identities appear naturally when evaluating Witten diagrams, as we demonstrate in the next subsection. Noting that a 4-momentum $\lambda \bar{\lambda}$ is invariant under the little group transformation $\left(\lambda,\bar{\lambda}\right)\rightarrow\left(\alpha\lambda,\alpha^{-1}\bar{\lambda}\right)$, we see that the above structures transform in the same way as an all-plus amplitude and will therefore be useful for simplifying that amplitude. It turns out that there are analogous identities for all helicities. The four cases are given by
\eqs{
\angb{n}{m}\angbb{n}{p}\angbb{n}{o} + \angb{m}{n}\angbb{m}{o}\angbb{m}{p} &= \angb{o}{p}\angbb{m}{o}\angbb{n}{o} + \angb{p}{o}\angbb{n}{p}\angbb{m}{p},\\
\ang{nm}\angbb{n}{p}\angbb{n}{o} + \angbb{m}{n}\angb{m}{o}\angb{m}{p}  &= \angb{o}{p}\angb{m}{o}\angbb{n}{o} + \angb{p}{o}\angbb{n}{p}\angb{m}{p},\\
\ang{nm}\angbb{n}{p}\ang{\bar{n}o} + \angbb{m}{n}\ang{mo}\angb{m}{p} &=
\angbb{o}{p}\ang{mo}\ang{\bar{n}o} + \ang{po}\angbb{n}{p}\angb{m}{p},\\
\ang{\bar{n}m}\angb{n}{p}\angb{n}{o}  +\ang{\bar{m}n}\angb{m}{o}\angb{m}{p} &= \angb{o}{p}\angb{m}{o}\angb{n}{o} +  \angb{p}{o}\angb{n}{p}\angb{m}{p},
\label{reflectionidentity}
}
where the four lines are all-plus, single-minus, alternating MHV and split MHV, respectively. The other cases can be obtained by conjugation. These identities are all consequences of momentum conservation and the Schouten identity, but proving them is slightly different for each helicity configuration. The identities in \eqref{reflectionidentity} can also be related to each other in different factorisation channels. For example, in the all-plus case we have
\eqs{
\angb{4}{1}\angbb{2}{4}\angbb{3}{4} + \angb{1}{4}\angbb{1}{3}\angbb{1}{2} = \angb{1}{2}\angbb{1}{3}\angbb{1}{4} + \angb{2}{1}\angbb{2}{4}\angbb{2}{3} -E^2\angbb{1}{3} \angbb{2}{4} \angbb{1}{2} \angbb{3}{4},
\label{eqn:MB in s and t}
}
which relates the $s$- and $t$-channel.

\subsection{Simplifying Witten Diagrams}
\label{app:BTerms}
In this subsection, we will explain how to simplify  the term $W_s^B$ in equation \eqref{eqn:sChannelBTerm} when written in terms of spinors, which is crucial for obtaining concise formulae for AdS$_4$ amplitudes. For concreteness, we will focus on the case $-+++$, since a similar strategy can be applied to other helicity configurations. Writing \eqref{eqn:sChannelBTerm} in spinor notation gives
\eqs{
W_s^{B,-+++}  &= \frac{1}{8k_1k_2k_3k_4}\frac{1}{Es} \Big(\angb{1}{2}^2\big(\angb{4}{3}\angbb{3}{4}(\angb{1}{4}\angbb{1}{4} - \angb{2}{4}\angbb{2}{4})\\
&\qquad\qquad + \angb{3}{4}\angbb{3}{4}(\angb{1}{3}\angbb{1}{3} - \angb{2}{3}\angbb{2}{3})\big)\\
&\qquad + \angbb{3}{4}^2\big(\ang{21}\angb{1}{2}(\angb{3}{2}\angbb{3}{2} + \angb{4}{2}\angbb{2}{4})\\
&\qquad\qquad + \angb{1}{2}\angbb{1}{2}(\ang{13}\angb{1}{3} + \ang{41}\angb{1}{4})\big)\\
&\qquad + 2\angbb{2}{4}^2\ang{12}\angb{1}{2}\angb{4}{3}\angbb{3}{4} + 2\angbb{2}{3}^2 \ang{12}\angb{1}{2}\angb{3}{4}\angbb{3}{4}\\
&\qquad - 2\angb{1}{3}^2\angb{1}{2}\angbb{1}{2}\angb{3}{4}\angbb{3}{4} - 2\angb{1}{4}^2\angb{1}{2}\angbb{1}{2}\angb{4}{3}\angbb{3}{4}\Big).
\label{wbmpppap}
}
Note that there are 4 terms with a factor of 2, and 8 terms without. These pair up in such a way that we can apply the Schouten a total of 8 times. Each of these gives a factor of $k_i$. For example, from the first line we have
\eqs{
\angb{1}{2}\angbb{3}{4}\angb{1}{4}\angb{4}{3}\big(\angb{1}{4}\angbb{2}{1} + \angb{1}{2}\angbb{1}{4}\big) &= -2ik_1 \angb{1}{2}\angbb{3}{4}\angb{1}{4}\angb{4}{3}\angbb{2}{4} ,\\
\angb{1}{2}\angbb{3}{4}\angbb{2}{4}\angb{4}{3}\big(\ang{21}\angbb{4}{2} + \angb{2}{4}\ang{\bar{2}1}\big) &= 2ik_2\angb{1}{2}\angbb{3}{4}\angbb{2}{4}\angb{4}{3}\angb{1}{4},
}
where we have added pieces from the last four terms and used the Schouten identity. We note that there is a factor of $\angb{1}{2}\angbb{3}{4}$ common to all terms. In addition, while there still seem to be a lot of unpromising terms, factoring out the $k_i$ gives
\eqs{
W_s^{B,-+++}  &= \frac{2i}{8k_1k_2k_3k_4}\frac{1}{Es} \Big((k_3 + k_4)\big(\angbb{1}{2}\angb{1}{3}\angb{1}{4} + \ang{12}\angbb{2}{4}\angbb{3}{2}\big),\\
&\qquad+(k_2 - k_1)\big(\angb{4}{3}\angb{1}{4}\angbb{2}{4} + \angb{3}{4}\angbb{2}{3}\angb{1}{3}\big)\Big).
}
We can then use the second line from \eqref{reflectionidentity} to get
\eq{
{W}^{B, -+++}_{s} = \frac{i \angb{1}{2}\angbb{3}{4}}{4k_1k_2k_3k_4} \frac{E - 2k_1}{Es}\big(\angbb{1}{2}\angb{1}{3}\angb{1}{4} + \ang{12}\angbb{2}{4}\angbb{3}{2}\big).
}

Similar steps can be applied for any helicity configuration, and the final result will be proportional to $E - 2\sum_{i\in-} k_i$. For completeness, we list the all-plus and MHV cases. Others can be obtained by relabelling or conjugation:
\eqs{
W^{B,++++}_s &= \frac{i\angbb{1}{2}\angbb{3}{4}}{4k_1k_2k_3k_4}\frac{1}{s}\big(\angb{2}{1}\angbb{2}{4}\angbb{3}{2} + \angb{1}{2}\angbb{1}{3}\angbb{4}{1}\big),\\
W^{B, -+-+}_s &= \frac{i\angb{1}{2}\angb{3}{4}}{4k_1k_2k_3k_4}\frac{E-2k_1 - 2k_3}{E s}\big(\ang{21}\angbb{2}{4}\angb{3}{2} + \angbb{1}{2}\ang{31}\angb{1}{4}\big),\\
W^{B, --++}_s &= \frac{i\ang{12}\angbb{3}{4}}{4k_1k_2k_3k_4}\frac{E-2k_1 - 2k_2}{E s}\big(\angb{1}{2}\angb{2}{4}\angb{2}{3} + \angb{2}{1}\angb{1}{3}\angb{1}{4}\big).
}

\bibliography{adsbcj}
\bibliographystyle{JHEP}

\end{document}